\documentclass[prl,twocolumn,showpacs,amsmath,amssymb]{revtex4}
\usepackage{graphicx}
\usepackage{latexsym}

\begin{document}
\title{The design of Kinetic Functionals for Many-Body Electron Systems : Combining analytical theory with Monte Carlo sampling of electronic configurations.}
\author{Luca M. Ghiringhelli and Luigi Delle Site}
\address{Max-Planck-Institute for Polymer Research, Ackermannweg 10, 55128 Mainz, Germany}

\date{\today}

\begin{abstract}
In a previous work [L.Delle Site, J.Phys.A {\bf 40}, 2787 (2007)]
the derivation of an analytic expression for the kinetic
functional of a many-body electron system has been proposed.
Though analytical, the formula is still non local
(multidimensional) and thus not ideal for numerical applications. In this work, by
treating the test case of a uniform gas of interacting spinless
electrons, we propose a computational protocol which combines the previous
analytic results with the Monte Carlo (MC) sampling of electronic
configurations in space. This, we show, leads to an internally
consistent scheme to design well founded local kinetic
functionals.
\end{abstract}


\maketitle

The design of electronic kinetic density functional is becoming a
field of rapidly growing interest due to the central role played
by this latter in Orbital-Free Density Functional (OFDFT) methods.
Such methods allow for large computational advantages compared to
standard Kohn-Sham (KS) based techniques since they are
computationally much less demanding
\cite{wang,watson,kaxiras,tfc,ortiz}. Moreover, as the
calculations are performed only in real space they allow for an
optimal design of an interface with classical codes to the aim of
connecting the electronic and the larger atomistic scale within a
unique computational framework \cite{ortiz2,ortiz3}. As discussed
above, the OFDFT is likely to play an important role in the
emerging multiscale modeling and simulation of condensed matter
systems, however the criterion to measure the validity of such an
approach is based on the quality of the kinetic functional
employed \cite{kaxiras2}. In principle, if we had the exact
kinetic functional the variational problem based on the
Hohenberg-Kohn theorem would not require the KS orbital-based
approximation anymore. In practice we must design kinetic
functionals, which are physically well founded and computationally
less demanding than the KS-based approaches. The most popular
currently used kinetic functional consists of the Thomas-Fermi and
Weizs\"{a}cker term, which are local, plus a linear response term
which is non local \cite{wang}. Recently one of the authors of the
current work has explored a different path than that above; this
is based on obtaining a kinetic functional for a system of $N$
electrons whose two central quantities are the one particle
electron density, $\rho({\bf r})$, and the ($N-1$)-conditional
probability density, $f({\bf r}_{2},...,{\bf r}_{N}|{\bf r}_{1})$
\cite{lui1,lui2}. The first is defined as: $\rho({\bf
r})=\rho({\bf r}_{1})=N\large\int_{\Omega_{N-1}}\psi^{*}({\bf
r}_{1},{\bf r}_{2},...{\bf r}_{N})\psi({\bf r}_{1},{\bf
r}_{2},...{\bf r}_{N})d{\bf r}_{2}...d{\bf r}_{N}$ (where
$\Omega_{N-1}$ is the $3(N-1)$ dimensional spatial domain of the
electrons and $\psi$ is the $3(N-1)$ dimensional wavefunction of
the electronic ground state) . The second term represents the
probability density of finding an $N-1$ electron configuration,
$({\bf r}_{2},...,{\bf r}_{N})$, for a given fixed value of ${\bf
r}_{1}$.  The function $f$ satisfies the following properties (see
also \cite{ayers}):
\begin{equation}
\large\int_{\Omega_{N-1}}f({\bf
  r}_{2},...,{\bf r}_{N}|{\bf r}_{1})d{\bf
  r}_{2}...d{\bf r}_{N}=1 \forall {\bf
  r}_{1}
\label{cond1}
\end{equation}
\begin{equation}
f({\bf
  r}_{1},...{\bf r}_{i}...{\bf r}_{j-1},{\bf r}_{j+1}...,{\bf r}_{N}|{\bf r}_{j})=0; \mathrm{for}~~ i=j;\forall i,j=1,N
\label{cond2}
\end{equation}
\begin{equation}
f({\bf
  r}_{1},...{\bf r}_{i}...,{\bf r}_{j}...{\bf r}_{k-1},{\bf r}_{k+1}...,{\bf r}_{N}|{\bf r}_{k})=0; \mathrm{for}~~ i=j;\forall i,j\neq k
\label{condeq3}
\end{equation}
In terms of $\rho$ and $f$ the equation to determine the
ground state of the systems writes \cite{lui1,lui2}:
\begin{widetext}
\begin{eqnarray}
E_{0} =
\min_{\rho}\left(\min_{f}\left(\Gamma[f,\rho]\right)+\frac{1}{8}\int\frac{|\nabla\rho({\bf
r}_{1})|^{2}}{\rho({\bf r}_{1})}d{\bf r}_{1} +\int v({\bf
r}_{1})\rho({\bf r}_{1})d{\bf r}_{1}\right) \label{eq4b}
\end{eqnarray}
where $v({\bf r}_{1})$ is the external potential and
\begin{eqnarray}
  \Gamma[f,\rho] = \int\rho({\bf
r}_{1})\left[\frac{1}{8} \int_{\Omega_{N-1}}\frac{|\nabla_{{\bf
r}_{1}}f({\bf r}_{2},....,{\bf r}_{N}|{\bf r}_{1})|^{2}}{f({\bf
r}_{2},.....,{\bf r}_{N}|{\bf r}_{1})}d{\bf r}_{2}....d{\bf
  r}_{N}+(N-1)\int_{\Omega_{N-1}}\frac{f({\bf r}_{2},.....,{\bf r}_{N}|{\bf r}_{1})}{|{\bf r}_{1}-{\bf r}_{2}|}d{\bf r}_{2}....d{\bf
  r}_{N}\right]d{\bf r}_{1}.
\label{eq5}
\end{eqnarray}
\end{widetext}
Eq.\ref{eq4b} is expressed in atomic units, i.e. $\hbar,m,e$ are
all equal to 1. This approach, up to this point is formally
rigorous; however, to proceed further it requires an explicit
expression for $f$. This latter is then built on the basis of the
mathematical prescription of
Eqs.\ref{cond1},\ref{cond2},\ref{condeq3} and a physical {\it
first principle} argument. The physical argument used in this case
is rather simple, though powerful, that is  in contrast to the
case of a gas of noninteracting electrons, the gas of interacting
electrons is characterized by the presence of the Coulomb
electrostatic term acting pairwise between electrons. The next
step consists of writing $f$ in terms of such an interaction in a
way that Eqs.\ref{cond1},\ref{cond2},\ref{condeq3} are satisfied
\cite{unif}. The proposed analytic form, consistent with the
requirements above, reads \cite{lui1,lui2}:
\begin{equation}
f({\bf r}_{2},...{\bf r}_{N}|{\bf r})=\Pi_{n=2,N}e^{{\overline{\overline E}}({\bf r})-\gamma E_{H}({\bf r},{\bf r}_{n})}\times \Pi_{i>j\neq 1}e^{-\gamma E_{H}({\bf r}_{i},{\bf r}_{j})}
\label{examp1}
\end{equation}
with the normalization condition:
\begin{equation}
e^{-{\overline{\overline E}}({\bf r})}=\int\Pi_{n=2,N}\Pi_{i>j\neq 1} e^{-\gamma E_{H}({\bf r},{\bf r}_{n})-\gamma E_{H}({\bf r}_{i},{\bf r}_{j})} d{\bf r}_{2}.....d{\bf r}_{N}
\label{examp1b}
\end{equation}
where $E_{H}({\bf r}_{i},{\bf r}_{j})=\frac{\rho({\bf
r}_{i})\rho({\bf r}_{j})}{|{\bf r}_{i}-{\bf r}_{j}|}$ and $\gamma$
is a free parameter to be optimized by minimizing
$\Gamma$ (w.r.t. $\gamma$) for a fixed $\rho$. The resulting
kinetic functional consists of a local term (the Weizs\"{a}cker term, the
second term on the r.h.s. of Eq.\ref{eq4b}) and an $N-$dimensional non local term known
as the {\it non-local Fisher Information} functional, which we will indicate as
$I[\rho]$ \cite{sears}(first term on the r.h.s. of Eq.\ref{eq5}).
The other term in $\Gamma$ is the Coulomb electron-electron term, which we will
indicate as
$C[\rho]$, and is also a multidimensional integral. The aim of
this work is to show, for a uniform gas how by employing a Monte
Carlo sampling of the electronic configurations in space, the non local
kinetic term, $I[\rho]$ (and in general the functional
$\Gamma[f,\rho]$) can be reduced {\it numerically} to a local
functional. Moreover, within the same approach, one can perform
numerically an {\it optimal} choice of the  $f$ by minimizing
$\Gamma$ w.r.t $\gamma$. The advantage w.r.t. previous methods is
that in this case the effects of the correlations between the
electrons regarding the kinetic functional are automatically
incorporated into a local functional. Moreover, within the OFDFT
approach, the optimization procedure for $f$ and the reduction of
the functional from non-local to local, can be also used  as an
intermediate step within a full self-consistent procedure as
suggested by Eq.\ref{eq4b}. This means that the kinetic functional for
an OFDFT algorithm can be also calculated on the fly without any
adjustable empirical parameters. This, as Eq.\ref{eq4b} shows, is
valid beyond the case of uniform gas that here we use as an
illustrative example \cite{spin}.

\begin{figure*}[t!]
\centering
\includegraphics[width=0.6\textwidth,clip]{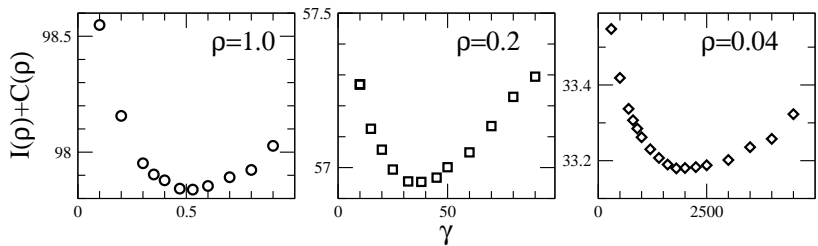}
\caption{The three panels show the values of $\gamma$ that
minimize the functional $I(\rho(\textbf{r}_1)) +
C(\rho(\textbf{r}_1))$ at three representative densities (for
N=100 electrons). Energies and densities are expressed in atomic
units.} \label{fig1.eps}
\end{figure*}

\begin{figure}[b!]
\centering
\includegraphics[width=0.58\columnwidth,clip]{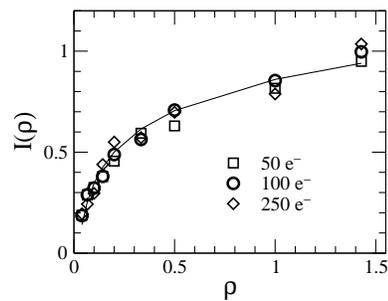}
\caption{Calculated values of $I(\rho(\textbf{r}_1))$ for N = 50, 100, and 250
electrons. The continuous line is the fit we propose (see text).}
\label{fig2.eps}
\end{figure}

{\bf MC evaluation of the non local part of the Kinetic
Functional:} At this point, the problem we must address to the aim
of obtaining a local functional from Eq. \ref{eq5} is that of
reducing the $3N$-dimensional integrals to $3$-dimensional ones.
The Metropolis MC approach has been shown to be ideal for this
kind of problems (see standard textbooks in the field,
e.g.\cite{mc1}, or Ref.\cite{cep1}). In general, the choice of
this kind of stochastic algorithms is a natural one for
high-dimensional problems, since their efficiency increases, if
compared to any other non-stochastic method, as the number of
dimensions of the integral increases. In this case, the
integration is done w.r.t. the variables ${\bf r}_{2},...{\bf
r}_{N}$ for a fixed value of ${\bf r}_{1}$ at a given density
$\rho({\bf r}_{1})$. By performing the integration for a large
number of fixed ${\bf r}_{1}$ values we then obtain the expression
of the functional local in ${\bf r}_{1}$. In order to evaluate Eq.
\ref{eq5} within the MC approach , we need to rewrite it as
follows:$ \Gamma[f,\rho] = \int\rho({\bf r}_{1}) \left[
I(\rho(\textbf{r}_1)) + C(\rho(\textbf{r}_1))\right] d{\bf
r}_{1}$, where the key quantities $I(\rho(\textbf{r}_1))$ and
$C(\rho(\textbf{r}_1))$ at a fixed ${\bf r}_{1}$ can be
conveniently written in the following form:
\begin{widetext}
\begin{eqnarray} \label{eq6}
I(\rho(\textbf{r}_1)) + C(\rho(\textbf{r}_1)) = \left[ \frac{1}{8}
\int_{\Omega_{N-1}} \mathrm{d}r^{N-1} f \ \left| \frac{
\vec{\nabla} f}{f} \right|^2 + \frac{1}{N} \sum_{i=1,N} \sum_{j>i}
\int_{\Omega_{N-1}}
 \mathrm{d}r^{N-1} \frac{f}{\left| \textbf{r}_i - \textbf{r}_j \right| } \right]_{{\rho(\textbf{r}_1)},\gamma}
\end{eqnarray}
\end{widetext}

In the above equation the two integrand functions are written as a
product of $f$, that acts here as a weighting function, and
another function (of all the coordinates). This is the form
suitable for the \emph{importance sampling MC} algorithm \cite{mc1}. In fact,
the integrals can be stochastically evaluated by using:

\begin{widetext}
\begin{equation}
I(\rho(\textbf{r}_1)) + C(\rho(\textbf{r}_1)) =\frac{1}{M}
\sum_{m=1,M} \left[ \frac{1}{8} \left| \frac{\vec{\nabla}_1 f}{f}
  \right|^2
  + \frac{1}{N} \sum_{i=1,N} \sum_{j>i} \frac {1}{| \textbf{r}_i - \textbf{r}_j |} \right]_{{\rho(\textbf{r}_1)},\gamma}
\label{mceq}
\end{equation}
\end{widetext}
where $M$ is the number of randomly chosen configurations, or
configurational space points, at which the function in the square
brackets is sampled. The sum in Eq. \ref{mceq} converges to the
integral of Eq. \ref{eq6} when the configurations are generated
according to $f$. In practice, since $f$ is not known explicitly
(the normalization factor is a multidimensional integral), a
Markov chain is constructed with limiting density $f$. For the
case of a uniform gas, the Metropolis MC algorithm consists of
randomly displacing the present configuration of the system of N
electrons and then accepting the new configuration, in a way that
the density of particles in average stays constant, with
probability (i.e. with acceptance rule):
\begin{eqnarray}
a(\mathrm{old} \rightarrow \mathrm{new}) = \min \left( 1,
\frac{f_{\mathrm{new}}}{f_{\mathrm{old}}} \right)
\end{eqnarray}
The robustness of the algorithm lies on the fact that the terms in the sum (Eq. \ref{mceq})
can be evaluated at each configuration with
relatively low computational cost. In fact, this is trivial for
the $1/| \textbf{r}_i - \textbf{r}_j |$ term, while some algebra
reduces the term $\left| \vec{\nabla}_1 f/f \right|^2$ as a
function of $\vec{\nabla}_1 E_H(\textbf{r}_1,\textbf{r}_n)$.

We modeled a uniform distribution of electrons by means of a
system of N electron (with N=10, 25, 50, 100, and 250) in a cubic
box. The MC scheme works as follows: one electron is selected at
random and a trial move is attempted. We adopt a trial move as a
uniformly distributed displacement of a randomly selected
electron. Explicitly, the acceptance rule is:
\begin{widetext}
\begin{equation} \label{MCmove}
    \frac{f_{\mathrm{new}}}{f_{\mathrm{old}}} =
    e^{- \gamma \left( E_H(\textbf{r}_1,\textbf{r}^\mathrm{new}_k) -
E_H(\textbf{r}_1,\textbf{r}^{\mathrm{old}}_k) \right)} \prod_{i
\neq 1,k} e^{- \beta
\left(E_H(\textbf{r}_i,\textbf{r}^\mathrm{new}_k) -
E_H(\textbf{r}_i,\textbf{r}^{\mathrm{old}}_k) \right)}
\end{equation}
\end{widetext}
where $k$ labels the attempted moved electron. If the move is
accepted the terms of the sum in Eq. \ref{mceq} are evaluated for
the average in the new configuration; otherwise another instance
of the quantities evaluated in the old configuration entries the
average. Periodic boundary conditions and minimum image convention
were imposed. In practice, each displaced electron was in turn the
center of the box and only one instance of each particle was used
in evaluating the relevant quantities in equations \ref{MCmove}
and \ref{mceq}. Periodic replicas of the system would be necessary
for the evaluation of the slowly decaying ``Coulomb'' dependence
of $C(\rho(\textbf{r}_1))$; in contrast, for the evaluation of
$I(\rho(\textbf{r}_{1}))$, it would be physically wrong to count
correlations of periodic replicas (therein considering also
self-correlations between the displaced electron and its images):
each electron should contribute once to the integral. For the
purposes of this paper, we did not need to worry about the finite
size dependence of $C(\rho(\textbf{r}_1))$, since we have notably
found out that both the value $\gamma$ that minimizes
$\Gamma[f,\rho]$ and $I(\rho(\textbf{r}_{1}))$ do not depend on the size $N$ of
the system.

For the case of uniform gas, once obtained the sum in Eq.
\ref{mceq}, the integral $\Gamma[f,\rho]$ is calculated by just
multiplying $I(\rho(\textbf{r}_1)) + C(\rho(\textbf{r}_1))$ by the
number of electrons N.

{\bf Numerical Results:} We performed the calculation at several
densities, selected in a way to resemble those of real metals. For
each density we evaluated the sum in Eq. \ref{mceq} at different
values of $\gamma$, searching for the value yielding the minimum
in the sum. Fig. \ref{fig1.eps}~shows the minimum found for
$\Gamma$ as a function of $\gamma$ for three different densities
(a) high, resembling the realistic case of Pt ($\sim
1.0~e/$bohr$^3$) (b) intermediate, resembling Ti ($\sim
0.2~e/$bohr$^3$), (c) low, resembling Na ($\sim 0.04~e/$bohr$^3$).
The existence of a minimum proves that indeed the problem is
variational and allows to find the optimal $\gamma$ that in this
way is no more an empirical parameter.
Having the optimal $\gamma$ for the different densities, we can
numerically express $I(\rho(\textbf{r}_{1}))$ and then fit the numerical results
to a functional form so that we can have an analytic expression.
The formula we obtain is: $I(\rho(\textbf{r}_{1}))=A+ B\ln(\rho(\textbf{r}_{1}))$, with $A=0.860
\pm 0.022$ and $B=0.224 \pm 0.012$.
The calculated values of $I(\rho(\textbf{r}_{1}))$ at the optimal
$\gamma$ are plotted in Fig. \ref{fig2.eps} together with the fit.
Finally, we can propose a new {\bf local} kinetic functional for
slowly varying densities, ready for OFDFT based codes:
\begin{equation}
K[\rho]=\frac{1}{8}\int\frac{|\nabla\rho({\bf
r}_{1})|^{2}}{\rho({\bf r}_{1})}d{\bf r}_{1}+ \int \rho({\bf r}_{1})\left( A+
B\ln\rho({\bf r}_{1}) \right) d{\bf r}_{1} \label{propf}.
\end{equation}

In conclusion, we have illustrated a computational protocol which
combines analytical and stochastic numerical approach to address
the problem of designing local kinetic functional for OFDFT based
codes. We have applied it to the test case of uniform interacting
gas of spinless electrons and given arguments from which it
emerges the general character of the approach for any system in
the ground state. Anyway, it must be said that despite being an
ideal system, the uniform gas has been often employed to develop
tools for realistic calculations (see e.g. the very recent work of
Drummond and Needs \cite{needs}), often under the approximation of
spinless particles. For this reason the specific results of this
work are already relevant beyond the role of illustrative example.
The approach shown here is internally (mathematically) consistent
and, computationally not expensive and rather flexible. It may
represent a direction along which to proceed since it can be
technically improved by both the mathematical and computational
point of view by improving the functional form of $f$ (as
discussed in Refs.\cite{unif,spin}) and the sampling of the
electronic configurations.

LMG acknowledges the AvH foundation for financial support. We
thank M.Praprotnik for a critical reading of the manuscript.


\begin{thebibliography}{99}
\bibitem{unif}
Of course, since among the properties required for $f$ there is
also that of mimicking the fermionic character, the {\it indirect}
interaction of the Pauli principle is also included. In case of
spinless particles, the Pauli principle is included in a crude
way, that is by the fact that two particles cannot have the same
state or position ${\bf r}$. In principle one may design an $f$
which can separate the interaction coming from the Pauli principle
and that from the Coulomb term. Here, for simplicity, we have used
only the Coulomb term since this form of $f$  does capture the
essential physics and it is numerically simple. However, in
principle it would not be difficult to include specific effects
(see also \cite{spin}).
\bibitem{cep1}
The general principles of the integration procedure are similar to
the ones employed (for example) for variational quantum MC
\cite{cep2}. However there are technical aspects which are
peculiar of our approach as discussed in the text.
\end{thebibliography}
\end{document}